\documentclass[10pt, two column, twoside]{IEEEtran}
\usepackage{color}
\usepackage[colorlinks, linkcolor=color1, anchorcolor=blue, citecolor=color1]{hyperref}

\usepackage{amssymb}
\usepackage{amsmath}
\usepackage[lined,boxed,commentsnumbered, ruled]{algorithm2e}
\usepackage{mathrsfs}
\usepackage{algorithmic}
\usepackage{bm}
\usepackage{hhline}
\usepackage{multirow}
\usepackage{pgfplots}
\usepackage{tikz}
\usetikzlibrary{arrows}
\usepackage{subfigure}
\usepackage{graphicx,booktabs,multirow}

\definecolor{colorhkust}{RGB}{20,43,140}
\definecolor{colortsinghua}{RGB}{116,52,129}
\definecolor{color1}{RGB}{128,0,0}

\date{}

\begin{document}

 \title{Towards Effective and Interpretable Semantic Communications}
\author{Youlong Wu, Yuanmin Shi, Shuai Ma, Chunxiao Jiang, Wei Zhang, and Khaled B. Letaief 
  \thanks{Y. Wu and Y. Shi  are with ShanghaiTech University. S. Ma is with Peng Cheng Laboratory.  C. Jiang is with Tsinghua University. W. Zhang is with The University of New South Wales.  K. B. Letaief is with The Hong Kong University of Science and Technology.  The corresponding author is Youlong Wu.}
 }

\maketitle

\maketitle

\begin{abstract}

 With the exponential surge in traffic data and the pressing need for ultra-low latency in emerging intelligence applications, it is envisioned that 6G networks will demand disruptive communication technologies to foster ubiquitous intelligence and succinctness within the human society. Semantic communication, a novel paradigm, holds the promise of significantly curtailing communication overhead and latency by transmitting only task-relevant information. Despite numerous efforts in both theoretical frameworks and practical implementations of semantic communications, a substantial theory-practice gap complicates the theoretical analysis and interpretation, particularly when employing black-box machine learning techniques. This article initially delves into information-theoretic metrics such as semantic entropy, semantic distortions, and semantic communication rate to characterize the information flow in semantic communications. Subsequently, it provides a guideline for implementing semantic communications to ensure both theoretical interpretability and communication effectiveness.

\end{abstract}

\section{Introduction}\label{Intro}


Conventional communication systems follow the design of Shannon's coding theory, which is agnostic of the meaning of information or task goal, and aims at accurately transmitting information bits over a noisy communication channel. In many applications, the goal of communication is to deliver information such that the destination can fulfill its task objective. Therefore, treating the information equally in terms of bits and delivering them without regard for the task's objective leads to redundant and unnecessary communication latency. Moreover, many emerging applications in 6G networks, such as autonomous driving, meta-universe, and Internet of Things, will involve massively connected intelligent machines with an exponential increase in traffic data, while simultaneously requiring ultra-low communication latency \cite{Shi'WC23}. It is envisioned that current communication systems will soon reach their limit, unable to sustain futuristic ubiquitous-connectivity and machine-intelligence services. This urgency prompts a fundamental reconsideration of communication system design, transitioning from conventional bit transmission to a focus on semantic or effectiveness levels. These two levels aim to address critical questions:  ``How precisely do the transmitted symbols convey the desired meaning?" and 
``How effectively does the received meaning affect conduct in the desired way?", respectively \cite{Weaver1953}.

 Semantic communication is a new communication paradigm concentrating on the semantic or effectiveness levels, involving the extraction and transmission of task-relevant information. 
 By selectively sending task-relevant information instead of the original source data, semantic communication substantially reduces the data traffic and achieves remarkably low end-to-end latency. It is envisioned that semantic communications will support efficient and timely services for real-time applications, such as autonomous driving, remote robotics, meta-universe, telehealth, etc \cite{Marios'CM}.  For instance, consider a remote water rescue robot that executes underwater (e.g., oceans, lakes, etc.) rescue missions. To ensure precise and prompt responses, the rescue robot  can exchange solely information pertinent to the rescued targets, instead of directly transmitting vast amounts of raw data collected from cameras and sensors.
 

 Due to its unparalleled advantages in communication efficiency, reliability, and its exceptional compatibility with emerging AI applications, semantic communication has garnered unprecedented interest in recent years. Extensive research efforts have been dedicated to various facets of this field, encompassing the definition of system measures and metrics (e.g., semantic entropy and semantic channel capacity) to quantify the amount of semantic information. Additionally, studies have focused on evaluating the performance of semantic communication from an information theory perspective \cite{Deniz'23}, as well as designing semantic communications for speech signals, images, language test, multimodel data, etc. 
 Despite many endeavors in theoretical frameworks and implementations of semantic communications, none of the existing theories had an operational impact. As a result,  current semantic communication systems lack unified theoretical guidelines, principles, and analytical tools. Here we identify four fundamental open problems: \begin{itemize}
 \item{Problem 1:} How can we accurately measure the quantity of semantic information embedded in the source and determine the effectiveness of semantic communication?  \item{Problem 2:} How to effectively implement semantic communications such that the destination(s) can make the right inference or action at the
right time? 

  \begin{table*}[t]
\begin{center}
\caption{A summary of semantic communications.}
\begin{tabular}{lcccccc}
\includegraphics[scale = 0.58]{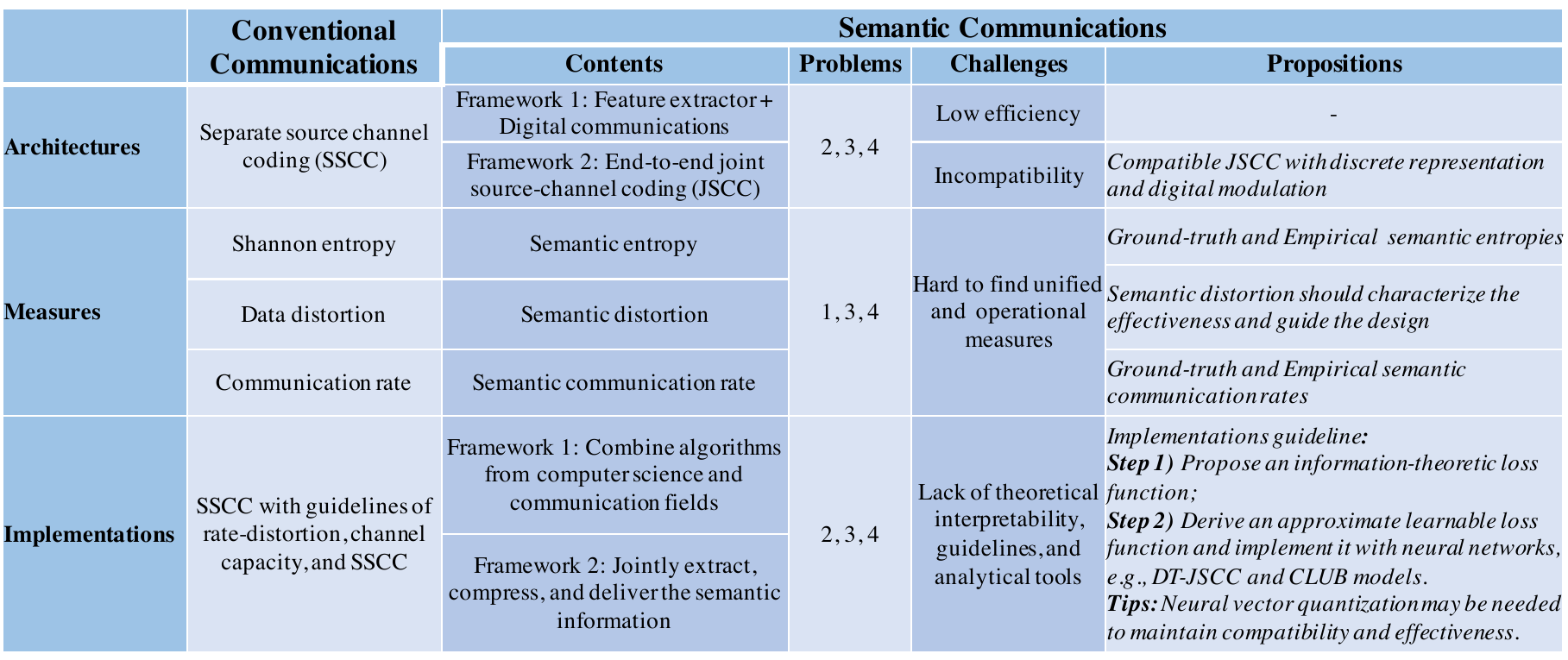}
 \label{tab:1}
\end{tabular}
\end{center}
\end{table*}

 \item{Problem 3:} How to \emph{optimally} extract and compress the semantic information, i.e.,  maximally compress the source while extracting all semantic information about the task objective?  
 \item{Problem 4:} How to \emph{optimally} deliver the semantic information in an end-to-end goal-oriented manner?  More precisely,  given a source and a task goal, what constitutes the sufficient and necessary condition, and how can it be achieved to successfully deliver semantic information that fulfills the task goal?
 
 \end{itemize}



 
 Problems 1 and 2 delve into semantic information metrics and semantic communication schemes, respectively. Problems 3 and 4 are two ultimate problems in semantic communications, holding a similar significance to establishing optimal source and channel coding in classical communications. Solving Problems 3 and 4 relies on solutions to Problems 1 and 2, but also necessitates addressing the crucial challenge of the converse proof, which validates optimality.

In this article, we mainly provide primitive solutions to Problems 1 and 2 from the perspective of information theory.    Specifically, for Problem 1, we introduce \emph{semantic entropy},  \emph{semantic distortion}, and \emph{semantic communication rate} to measure the quantity of semantic information, the distortion of semantic information recovery, and the effectiveness of semantic communication, respectively. Regarding  Problem 2, we propose a comprehensive guideline for designing semantic communications with \emph{theoretical interpretability} and \emph{communication effectiveness}, and illustrate how to apply the guideline to design robust, privacy-preserving, and compatible semantic communications.   {Our goal is to provide theoretical insights and methodologies in the field of semantic communications, with the expectation that the corresponding results could shed light on solutions to Problems 3 and 4.} In Table \ref{tab:1}, we summarize the frameworks, challenges, and our propositions in semantic communications.

 The rest of this article can be outlined as follows. Section \ref{SecArchitectures} presents the general architectures for semantic communications. Section \ref{Measures} provides three types of important measures including semantic entropy, semantic distortion, and semantic communication rate. Section \ref{Implementations} demonstrates how to use information theoretical tools and measures to design interpretable and efficient semantic communication. Finally, we conclude this article in Section \ref{SecConclusion}. 
 

\section{Architecture}\label{SecArchitectures}

This section outlines the modules and frameworks necessary to characterize the information flow within semantic communication systems. 

 \begin{figure*}[t] \centering
\includegraphics[scale = 0.55]{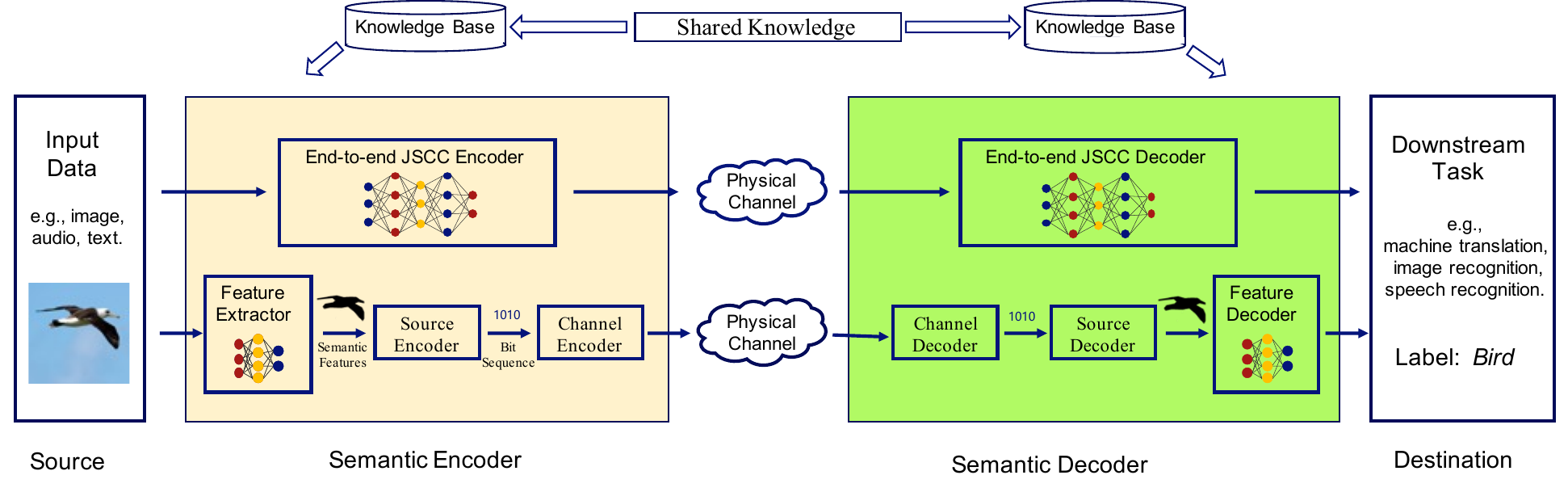}
\caption{An architecture of task-oriented semantic communication systems.}
 \label{Fig:architecture}
\end{figure*}

\subsection{Modules}

A semantic communication framework generally contains the following modules: Information source, knowledge base, semantic encoder, physical channel, and semantic decoder. 
\begin{itemize}
\item{Source}: The source produces source samples such as text, speech, image, video, etc. Given the task goal, the source samples carry task-relevant information desired by the destination. Generally, the task-relevant information is intrinsically hidden in the source, and cannot be directly observed by the source. 

\item{Knowledge Base}: 	The knowledge base contains all the necessary information that can facilitate semantic communications at the semantic level, such as the training
dataset and background knowledge. 

 	\item{Semantic Encoder}:
		Based on the knowledge base, the semantic encoder extracts the task-relevant information and encodes it into signals that will be transmitted over the channel. When the physical channel is noiseless or when focusing on semantic source coding, the semantic encoder plays the role of semantic information extraction and compression. To communicate semantic information over the noisy channel, the semantic encoder will also play the role of channel coding, i.e., adding coding redundancy to improve the communication reliability.

 	\item{Physical Channel}:
	The physical channel is the physical medium between the transmitter and destination, and is the same for conventional communication systems.

\item{Semantic Decoder}:
	Based on the received channel output and the local knowledge base, the semantic decoder aims to recover the semantic information with a prescribed distortion to support the downstream task. 
		
		\end{itemize}
		
 Given a semantic communication system,  characterizing the semantic information flow running through each module requires information measures and metrics, such as semantic entropy, semantic distortion, and semantic communication rates. These will be formally defined in Section 	\ref{Measures} from the perspective of information theory.
		
		 \subsection{Frameworks}

In general, there are two types of frameworks for implementing semantic communication, as shown in Fig. \ref{Fig:architecture}: one type separately designs semantic source coding and channel coding, and the other type designs the semantic encoder in an end-to-end JSCC manner. In the first type, the task-relevant semantic information hidden in the source data is extracted and compressed via a feature extractor, followed by digital communication which converts the semantic information into bits and sends them via digital modulation. 
The second type applies an end-to-end JSCC strategy that directly maps the source data to channel symbols without the binary interface. 

The first type is easy to implement but it does transmit the semantic information with lower efficiency. In contrast, the second type potentially achieves higher communication effectiveness, but often suffers from the compatibility issue. Specifically, in the first framework,  the semantic source coding and digital communications are separately designed, allowing for a simple fusion of computer science and communication technologies. However,  this separate design could lead to suboptimal   performance since the semantic encoder completely ignores communication uncertainties  In the second framework, the extraction, compression, and delivery of semantic information are jointly designed via end-to-end JSCC strategy, which outperforms separate approaches in low-latency and semantic-level transmission \cite{Bourtsoulatze2019DeepJS}. The end-to-end JSCC strategy, however, encounters compatibility challenges in modern digital communication systems that ignore the semantic meaning and adopt an SSCC strategy.  In Section \ref{Implementations}, we will introduce a guideline for designing the second type of framework, and illustrate how to design JSCC communication with theoretical interpretability,  heightened communication effectiveness, and improved compatibility.

\section{Interpretability: Measures}\label{Measures}
      In semantic communication frameworks, three key measures are crucial: \emph{semantic entropy, quantifying the semantic information in a source; \emph{semantic distortion}, measuring the difference between desired and recovered semantic information; and \emph{semantic communication rate}, indicating the semantic information received by the destination.
  Despite numerous attempts, a consensus on performance measures for semantic communications remains elusive \cite{Deniz'23}.} In particular, the existing definitions often lack the operational impact on the implementations of semantic communication. In this section, we ask a reverse question: Are the measures of Shannon information theory not capable to characterize the performance of semantic communication? We attempt to answer this question in a positive way.  
 
\subsection{Semantic Entropy}\label{SecSE}
 
 Semantic entropy measures the quantity of semantic information embedded in the source.
  In the following, we introduce universal definitions of ground-truth and empirical semantic entropy based on the measure of Shannon entropy. 

\subsubsection{Ground-truth Semantic Entropy}

Given the source data and task objective, the most compact task-relevant information, referred to as the \emph{ground-truth semantic information} (GSE), is a minimal sufficient statistic (MSS) of the source for the task goal, which should have invariant Shannon entropy. Recall that Shannon entropy is a universal measure that quantifies the uncertainty (or the amount of bit information) contained by a random variable (RV), and is only determined by its stochastic probability distribution. If the stochastic distribution of ground-truth semantic information is known, the Shannon entropy is sufficient to measure the amount of semantic information. Thus, given a task-oriented communication, we define the GSE as the Shannon entropy of the MSS of the source data for the communication goal. The GSE is computable when the stochastic distribution of the semantic information is known in advance or can be perfectly obtained via an ideal semantic source encoder (i.e., a semantic source encoder that can extract the semantic information about the communication goal). For example, in the remote classification task where a transmitter already knows the label of the source image, and wishes to tell the destination about the label information, then the semantic entropy is simply  Shannon entropy of the label information.

\subsubsection{Empirical Semantic Entropy}
In the general case, the stochastic distribution of ground-truth semantic information with respect to the communication goal is not known priorly.  This is primarily due to the high complexity of the task or the insufficient computation capability of the semantic encoder. Consequently, direct calculation of GSE becomes unfeasible. To address this challenge, we introduce the concept of empirical semantic entropy (ESE) for arbitrary semantic communication systems. It is defined as the Shannon entropy of the extracted representations (or features) from the semantic source encoder without injecting coded redundancy to combat channel uncertainty, i.e., ignoring the noise and fading of the physical channel.   



Three aspects of this definition should be highlighted: i) Unlike the GSE which only depends on the source and task goal (irrelevant to the semantic source encoder), the ESE is related to the design of the semantic source encoder. When the semantic source encoder is optimal, then the ESE equals the GSE; ii) The channel variation is ignored. This is to ensure that the feature extractor only aims to extract and compress the stochastic distribution of semantic information. Otherwise, the semantic source encoder will additionally play the role of channel coding that adds redundancy to combat channel uncertainty; iii) The semantic source encoder may not be optimal, i.e., it may not extract the MSS of the source about the task goal. In this way, the ESE can be larger or smaller than the GSE. A smaller ESE indicates semantic loss in the extracted semantic feature, consequently impairing the utility performance of semantic communication. Conversely, when the ESE surpasses the GSE, the extracted information contains task-irrelevant data that may result in redundant communication overhead.


\subsection{Semantic Distortion}
Semantic distortion is a measure of the cost of representing the semantic information by the recovered semantic information. If the ground-truth semantic information is known and the distortion measure is given, then the semantic communication problem can be viewed as a classical source compression problem and has been extensively investigated by information and coding theory. Unfortunately, acquiring ground-truth semantic information is typically arduous, which makes the computation of distortion between the ground-truth and reconstructed semantic information hard to compute. Even if access to ground-truth semantic information is feasible, finding appropriate distortion measures remains challenging due to the diversity or complexity of tasks.  
To address this issue, we propose that \emph{a well-defined semantic distortion measure should be able to characterize the effectiveness of information reconstructions, and also explicitly guide the compression and transmission of semantic information.} In the following, we introduce two of such measures including   {classification distortion} and  {perception distortion}.


In classification tasks where the destination wishes to recognize the categories of the data transferred by the transmitter,  the most effective performance metric is the classification error (or accuracy).
Despite its significance, classification error is probably not an ideal semantic distortion measure, as it is hard to build the theoretical connections between the classification error with the validity of the extracted semantic information. Instead, we propose defining a classification distortion as the Shannon entropy or cross entropy of the category label conditional on the observations at the destination. The cross entropy,   widely applied in the loss function design of machine learning algorithms, has exhibited considerable success in classification tasks. 

In generative tasks, the destination wishes to generate source symbols that look like a real data sample in the data set or look natural from the human viewpoint. 
This inspires the investigation of perceptual naturalness metrics in different tasks. Blau and Michaeli investigated the perceptual naturalness and defined perception distortion as the distance (e.g., Kullback-Leibler divergence) between the probability distributions of restored data and the desired probability distribution \cite{Blau2018ThePT}. The perception distortion can be found in the generative learning algorithms, such as the generative adversarial network (GAN) \cite{Goodfellow2014GenerativeAN}, which provides empirical evidence of applying perception distortion to practical AI applications.   Recently, the trade-off between perception, distortion, and classification have been investigated from both theoretical analysis and experimental results.  

 \subsection{Semantic Communication Rate}
The semantic communication rate measures the amount of semantic information obtained by the destination in semantic communication. In essence, the effectiveness of semantic communications depends on the task goal. If the semantic information is perfectly known and the distortion measures are given, then the semantic communication falls into the realm of a JSCC problem. Unfortunately,  the ground-truth semantic information and appropriate distortion both are hard to obtain in most cases. 
In this subsection, we propose two universal metrics: Ground-truth and empirical semantic communication rates,  to characterize the effectiveness of semantic communications with known and unknown ground-truth semantic information,  respectively.


 

In information theory, the rate of JSCC is defined as the average number of symbols communicated per channel use. The achievable communication rate, measured in bits per channel use,  equals the product of the rate of JSCC and the rate-distortion function   \cite{Gamal2011NetworkIT}. Following a similar definition, we define the \emph{ground-truth semantic communication rate} (in bits per channel use) as the product of the average number of semantic source symbols communicated per channel use and the mutual information between the ground-truth semantic information and recovered semantic information. For example, if in each channel slot, one semantic source symbol can be perfectly delivered to the destination,  (i.e., the rate of JSCC is 1 and the ground-truth semantic information is perfectly recovered by the destination), 
then the semantic communication rate equals the GSE. 

Due to the obscurity of the ground-truth semantic information in most cases, computing the semantic communication rate becomes challenging. Alternatively, we define the \emph{emprical semantic communicate rate} as the product of the average number of communicated feature symbols per channel use and the mutual information between the extracted semantic information and recovered semantic information. If in each channel slot, one semantic source symbol can be perfectly delivered to the destination, 
then the semantic communication rate equals the empirical semantic information semantic entropy. If the recovered semantic features are not exactly the same as the empirical semantic information, then there exists semantic channel noise incurred by communication uncertainties such as channel fading and noise.

The ground-truth semantic communication rate measures the quantity of the ground-truth semantic information successfully delivered by the transmitter in each channel transmission. It can serve as a performance metric that evaluates the performance of semantic communication if it is computable. In other words, a higher ground-truth semantic communication rate implies better communication performance. On the other hand,  the empirical semantic communication rate may not directly evaluate the exact performance of semantic communication. This is because the extracted information from the semantic source encoder could encompass task-irrelevant information or only capture a portion of task-relevant information.  In essence, the empirical semantic communication rate is a metric that characterizes how communication uncertainties (e.g., channel fading/noise and quantization error)   impact the communication effectiveness of the extracted semantic information, not necessarily the ground-truth semantic information.

 It is worth mentioning that calculating semantic entropy, semantic distortion, and semantic communication rate requires sophisticated estimations of probability distributions,  entropy, and mutual information for high-dimensional data, which have been extensively explored in the literature. Typically, the estimation involves two types of approaches:  nonparametric techniques like kernel density estimators and nearest neighbors, 
 and parametric methods that model probability distributions with specific functional forms governed by several parameters, e.g., the variational auto-encoder produces output with a Gaussian distribution \cite{Kingma2013AutoEncodingVB}. 

\begin{figure*}[h] \centering
\includegraphics[scale = 0.46]{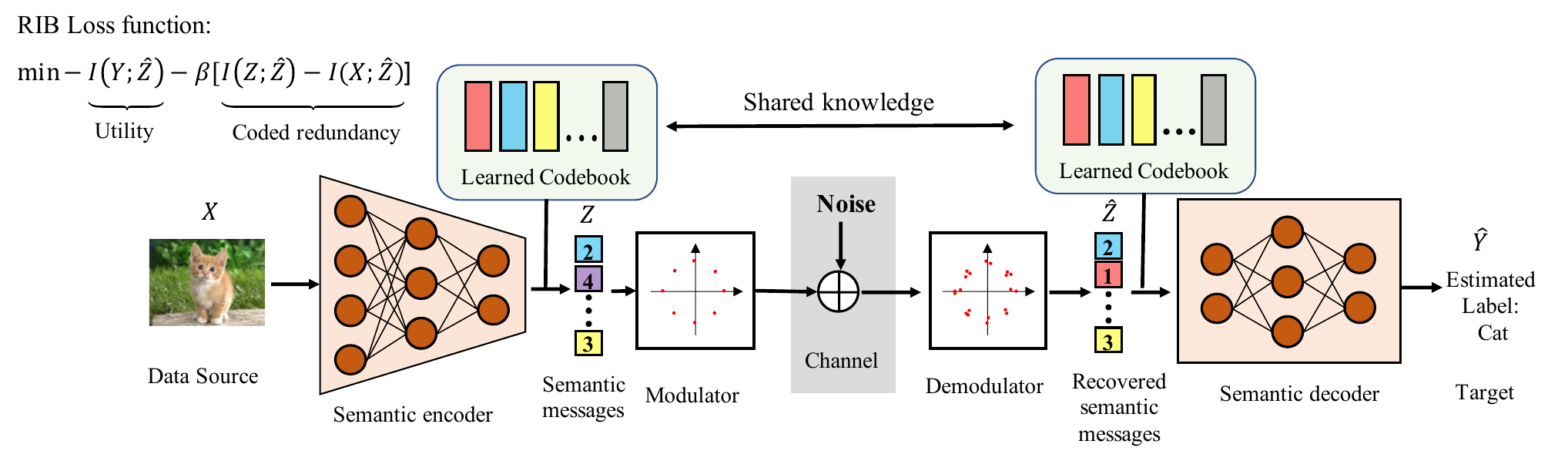}
\caption{The RIB framework for semantic communication with discrete modulations.}
 \label{Fig:RIB}
\end{figure*}

\section{Effectiveness of Semantic Communications: Implementations}\label{Implementations}

 A well-designed semantic communication strategy should embody three key merits: Theoretical interpretability, guidable principle, and competitive effectiveness. Besides,  high compatibility with existing communication systems would be another merit from the perspective of practical use. In the following, we first present a general guideline on utilizing information theoretical tools and measures proposed in Section \ref{Measures} to design interpretable and effective JSCC semantic communications.We then provide two illustrative examples demonstrating how to design practical semantic communications that ensure robustness and privacy guarantees, respectively.
 
\subsection{A Design Guideline for Semantic Communication}\label{SecGuidline}

Semantic communication typically involves diversified task goals. To establish interpretable and efficient semantic communications, we propose a guideline based on the architectures and semantic distortion measures described in Sections II and III, respectively. The guideline in general contains two steps: 

\begin{enumerate}
\item Design an information-theoretic loss function according to the semantic communication requirements (e.g., preserving robustness against channel variations or preserving privacy to protect sensitive information) with information-theoretic metrics such as Shannon entropy, semantic entropy, mutual information, and semantic communication rate; 
\item Derive a learnable loss function by finding variational approximations of the information-theoretic loss functions, and then train deep neural networks (DNNs) according to the learnable loss function. To ensure compatibility with digital modulation, neural vector quantization could be employed after or embedded in the DNN to maintain the well-learned information structure of semantic representation.  
\end{enumerate}



 The first step is instructive and furnishes an interpretable framework for the learning process, whereas the second step is the most challenging, primarily because of the intricacies involved in computing mutual information in DNNs.
The variational information bottleneck (VIB) method, introduced in \cite{DeepVI}, offers a promising solution to tackle this challenge. Firstly, it establishes a variational bound for information-theoretic loss functions, followed by parameterizing the encoding and decoding distributions using a family of distributions whose parameters are determined by DNNs. The distributions are then estimated by the Monte-Carlo sampling. Finally, the optimization problem is solved via a reparameterization trick \cite{Kingma2013AutoEncodingVB} and stochastic gradient descent algorithms.  Next, we present two examples to illustrate how to use this guideline to design semantic communication with robustness against channel variations and privacy guarantees.

\subsection{Robust Information Bottleneck for Semantic Communications}

 The original information bottleneck rule (IB) is to characterize the rate-distortion tradeoff between the transmission load and the achievable inference accuracy of the extracted features \cite{Tishby2000TheIB}. It extracts the most relevant information about the ground-truth semantic information while filtering the irrelevant information as much as possible. This is achieved by maximizing the mutual information between the inference result and the extracted semantic information, while minimizing the mutual information between the input data and the extracted representation.   The IB-based methods have been widely applied to reduce dimensionality to ensure high accuracy, generative abilities, and robustness \cite{abdelaleem2024deep}.  Note that the original IB rule only focuses on the compression and extraction of semantic information, ignoring the transmission over the noisy channel. A recent work by Shao \emph{et al.} has extended IB to semantic communication with channel noise and proposed a variational feature encoding  (VFE) framework to extract and deliver semantic information \cite{Shao2022LearningTC}. 
  However, this method still suffers from severe channel uncertainty, mainly because   1) In the low signal-to-noise ratio (SNR) regime, the continuous values of extracted representations need to be quantized for later transmission, while the quantization will destroy the learned information structure of semantic information; 2) The effect of the channel noise is not fully considered in their loss function design, i.e., coding redundancy added in the semantic encoder is not sufficient to combat channel noise.
 
\begin{figure}[t]
 \centering
  \includegraphics[scale = 0.5]{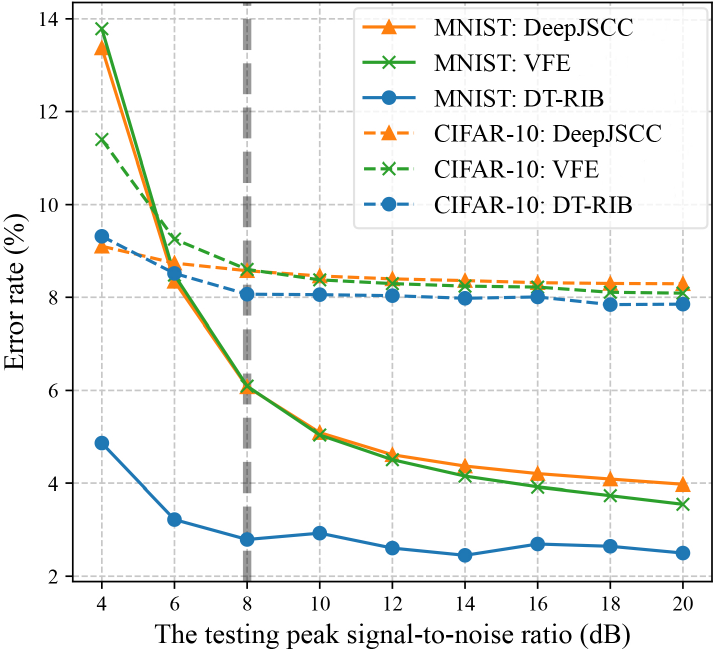}
\caption{Performance of the proposed DT-RIB, DeepJSCC \cite{Bourtsoulatze2019DeepJS}, and VFE \cite{Shao2022LearningTC} on MNIST and CIFAR-10 dataset with  8dB training  peak SNR.}
  \label{Fig:Simulations}
 \end{figure}

 To address the problem, we propose a robust Information bottleneck framework with digital modulation  (DT-RIB) that exhibits both good effectiveness and compatibility \cite{Wu'JSAC},  see Fig. \ref{Fig:RIB}. Following the guideline in Section \ref{SecGuidline}, we first design an information-theoretic loss function that maximizes the mutual information between the channel output and the ground-truth semantic information  (i.e., maximizing the information utility), while simultaneously maximizing the coded redundancy 
 to improve the robustness against channel noise.  In the second step, we derive a variational upper bound for the information-theoretic loss function,  and then apply discrete representation encoding to obtain \emph{discrete} semantic messages, followed by digital modulation to deliver semantic messages. More specifically, we first learn a discrete task-related codebook that is dependent on the dataset, and then map each input data into a subset of codewords, whose indices are the output of the semantic encoder and are referred to as semantic messages.  Subsequently, the semantic messages are directly modulated (e.g., using 16-PSK modulation) and sent. To ensure the differentiability and trainability of the probabilistic model using the backpropagation algorithm, we resort to continuous reparameterization based on the Gumbel-softmax distribution. Fig. \ref{Fig:Simulations} presents the quantitative comparison of the inference performance of MNIST and CIFAR-10 datasets  under various testing peak  signal-to-noise ratio (SNR) with 8dB training peak   SNR. In the evaluation, we consider DeepJSCC \cite{Bourtsoulatze2019DeepJS} and VFE \cite{Shao2022LearningTC} as baselines with the same DNN architecture as ours.   Given that both DeepJSCC \cite{Bourtsoulatze2019DeepJS} and VFE are incompatible with digital modulations, we assess them using analog modulations, thereby avoiding performance degradation through quantization.   It can be seen that when the proposed DT-RIB   achieves fewer classification errors than the baselines, and exhibits more robustness against channel fluctuations.    To adapt the learning network to the dynamic channel variations, SNR information could be incorporated by attention mechanisms such that the learned features can adaptively fit to the channel conditions.   

 \begin{figure*}[t] \centering
\includegraphics[scale = 0.50]{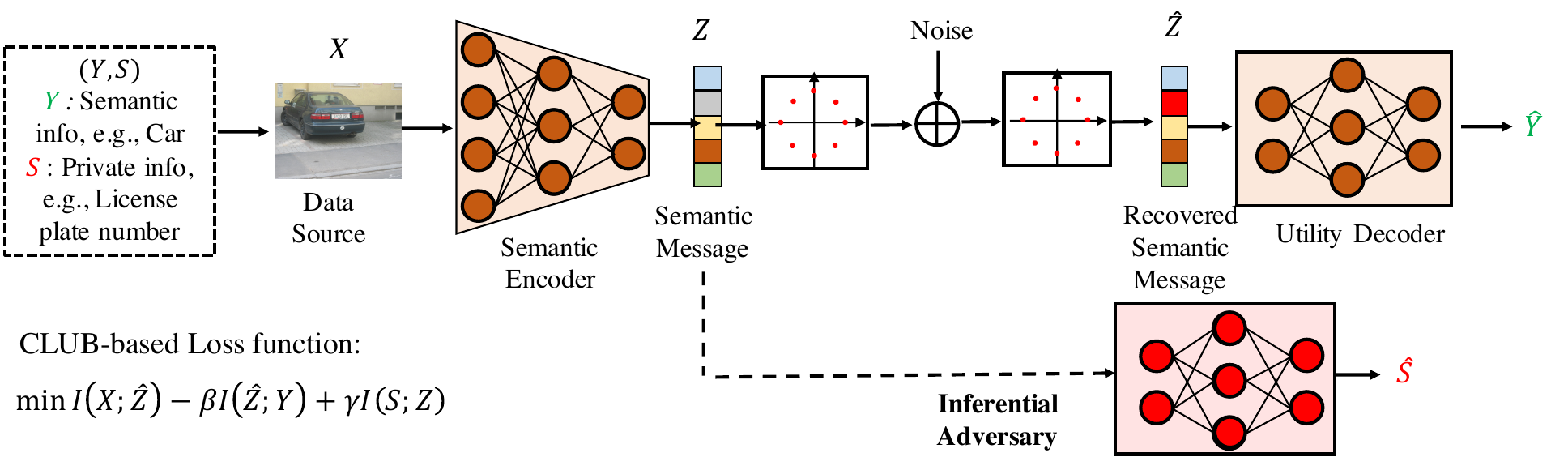}
\caption{The CLUB-based JSCC for privacy-preservative semantic communications.}
 \label{Fig:CLUB}
\end{figure*}

\subsection{Privacy-preserving Semantic Communications}

In privacy-preserving semantic communication, the transmitter only wishes to send task-relevant information to the destinations while simultaneously keeping private information unknown to the destination. For example, a self-driving car may wish to detect neighboring cars and pedestrians from camera video while not willing to extract privacy-sensitive info such as license plate numbers pedestrians' faces. To extract the semantic representation with guarantees of both utility and privacy, the complexity-leakage-utility bottleneck (CLUB) model \cite{Razeghi2022} has been proposed. It maximizes the information utility (i.e., the mutual information between the extracted representation and the ground-truth semantic information), while simultaneously 
minimizing information complexity (i.e., the mutual information between the extracted semantic representation and the source data), and minimizing the privacy leakage (i.e., the mutual information between the extracted representation and the private semantic information). Based on the CLUB model, variational bounds on the information utility, complexity, and privacy-leakage have been derived, respectively, which jointly constitute a learnable inference bound. The learnable inference bound is then parameterized via DNN including a semantic encoder network,  a utility decoder network, and an adversary inferential decoder network, as shown Fig. \ref{Fig:CLUB}.

The CLUB model simply focuses on the extraction and compression of the semantic information with a privacy guarantee, while ignoring the transmission over noisy channels. To implement the CLUB model over noisy channels, a simple approach is to apply the first framework in Fig. \ref{Fig:architecture}, where the CLUB learning model is adopted to extract and compress the semantic representation, followed by digital communication to quantize and deliver the extracted representation. This is easy to implement and is compatible with the existing communications systems, but it leads to low efficiency because the quantizer destroys the well-learned structure and channel noise may severely degrade utility performance. Alternatively, we could apply the JSCC with discrete representations and digital modulations as in the DT-RIB model. More specifically, the semantic encoder  maximizes the information utility 
(i.e., the mutual information between the \emph{recovered} representation   and the ground-truth semantic information), while simultaneously 
minimizing the information complexity (i.e., the mutual information between the \emph{recovered} semantic representation and the source data) and minimizing the privacy leakage for the worst case where an inferential adversary can obtain perfect (i.e., the mutual information between the \emph{extracted} representation and the private semantic information). 
 Then, the transmitter maps the source data to semantic messages represented by codeword indices, delivering them through digital modulations. The introduction of stochasticity (e.g., channel noise) serves as an implicit regularization technique for the encoder, aiding the encoding process in learning a meaningful latent space and enhancing adversarial robustness.   Therefore, the  CLUB-based JSCC strategy could facilitate private, efficient, and robust communications, while maintaining a certain level of compatibility with existing digital communication systems.

\section{Conclusion}\label{SecConclusion}

Semantic communication, which focuses on communications at the semantic or effectiveness level, is a disruptive paradigm for future communication systems  such as automobile transportation, meta-universe, telehealth, etc.  To establish interpretable, effective, and reliable semantic communications with theoretical analysis and guidelines, information theory will play important roles in finding performance measures and metrics, guiding feasible and interpretable scheme design, as well as providing analytical tools to reveal intrinsic relations among utility, privacy, complexity, etc. However, due to the requirements on stochastic distributions and the difficulties in computing information measures (e.g., mutual information) in neural networks, designing a reliable, low-latency, and high-utility semantic communication requires bridging machine-learning and information-theory communities to jointly consider the extraction, compression, and transmission of semantic information. 
 Although only typical architectures, measures, implementations, and examples were presented, we hope that this article will inspire more exciting theories (e.g., establishing the optimal semantic source coding and channel coding theorems), implementations, and applications for semantic communications.
 
 \section{Acknowledgments}
This work was supported in part by the National Natural Science Foundation of China under Grant Nos.  61901267, 62271318, 62325108, 62341131, the Hong Kong Research Grants Council under the Areas of Excellence scheme grant AoE/E-601/22-R, the Natural Science Foundation of Shanghai under Grant No. 21ZR1442700, Guangdong Basic and Applied Basic Research Foundation under grant  2024A1515030028, and the Shanghai Rising-Star Program under Grant No. 22QA1406100. 

\bibliography{TowardsSemantic}
\bibliographystyle{ieeetr}

\vspace{3mm}
{

{\bf{Youlong Wu}} [S'11-M'14] (wuyl1@shanghaitech.edu.cn) received the  B.S. degree from Wuhan University and the Ph.D. degree from Telecom ParisTech. He is now an Associate Professor at the School of Information Science and Technology, ShanghaiTech University. He received the TUM Fellowship in 2014 and is an Alexander von Humboldt Research Fellow.\\

{\bf{Yuanming Shi}} [S’13-M’15-SM’20] (shiym@shanghaitech.edu.cn)   received the B.S. degree from Tsinghua University and the Ph.D. degree from The Hong Kong University of Science and Technology. Since September 2015, he has been with the School of Information Science and Technology in ShanghaiTech University, where he is a Full Professor. He is a recipient of the IEEE Marconi Prize Paper Award in Wireless Communications in 2016, the Young Author Best Paper Award by the IEEE Signal Processing Society in 2016, the IEEE ComSoc Asia-Pacific Outstanding Young Researcher Award in 2021, and the Chinese Institute of Electronics First Prize in Natural Science in 2022, and the Best Paper Award from IEEE International Mediterranean Conference on Communications and Networking (MeditCom) in 2023.  He is an IET Fellow. \\

{\bf{Shuai Ma}} [S'13-M'16] (e-mail: mash01@pcl.ac.cn) received the B.S. and Ph.D. degrees in communication and information systems from Xidian University, Xi'an, China, in 2009 and 2016, respectively. From 2016 to 2019, he has been an associate Professor with the School of Information and Control Engineering, at the China University of Mining and Technology, Xuzhou, China. He is now an Associate Researcher at Peng Cheng Laboratory, Shenzhen, China. \\

 {\bf{Chunxiao Jiang}} [S’09-M’13-SM’15-F'24] (jchx@tsinghua.edu.cn) is an associate professor in the School of Information Science and Technology, Tsinghua University. He received a B.S. degree in information engineering from Beihang University, Beijing, in 2008, and a Ph.D. degree in electronic engineering from Tsinghua University, Beijing, in 2013, both with the highest honors. \\
 
 {\bf{Wei Zhang}} [S’01-M’06-SM’11-F’15] (w.zhang@unsw.edu.au) received the Ph.D. degree in electronic engineering from the Chinese University of Hong Kong in 2005. He is currently a Professor with the University of New South Wales, Australia. His current research interests include 6G communications and networks. \\

 {{\bf{Khaled B. Letaief}} [S'85-M'86-SM'97-F'03] (eekhaled@ust.hk) received his Ph.D. degree from Purdue University. He has been with HKUST since 1993, where he was Acting Provost and Dean of Engineering, and is now a Chair Professor and the New Bright Professor of Engineering. 
From 2015 to 2018, he joined HBKU in Qatar as Provost. He is an ISI Highly
Cited Researcher and a recipient of many distinguished awards. He has served
in many IEEE leadership positions including ComSoc President,
Vice-President for Technical Activities, and Vice-President for Conferences. 
He is a Member of US National Academy of Engineering.\\

}

\end{document}